\documentclass[11pt]{article}
\usepackage{latexsym}
\usepackage{amssymb,cite,epsfig}

\def\be{\begin{equation}}
\def\ee{\end{equation}}
\def\bea{\begin{eqnarray}}
\def\eea{\end{eqnarray}}
\def\ba{\begin{array}}
\def\ea{\end{array}}
\def\nn{\nonumber \\}

\newcommand{\ul}{\underline}

\renewcommand{\a}{\alpha}
\renewcommand{\b}{\beta}

\renewcommand{\d}{\delta}
\newcommand{\Gphi}{\Gamma^{\ul{\varphi}}}
\newcommand{\hGphi}{\hat\Gamma^{\ul{\varphi}}}

\begin{document}
\begin{flushright}
IFUM-760-FT\\
\end{flushright}
\vspace{.5truecm}

\centerline{\Huge  Actions and Fermionic symmetries }
\centerline{\Huge  for D-branes in bosonic backgrounds }
\vspace{1truecm}

\begin{center}
\renewcommand{\thefootnote}{\fnsymbol{footnote}}
 {\Large Donald~Marolf$^{1}$\footnote{marolf@physics.syr.edu},
 Luca~Martucci$^{2,3}$\footnote{luca.martucci@mi.infn.it}
 and Pedro~J.~Silva$^{2,3}$\footnote{pedro.silva@mi.infn.it}}\\
\renewcommand{\thefootnote}{\arabic{footnote}}
\setcounter{footnote}{0}
\vspace{.5truecm} {\small \it $^1$ Physics Department, Syracuse
University,\\ Syracuse, New York, 13244, United
States

\vspace*{0.5cm}

$^2$ Dipartimento di Fisica dell'Universit\`a di Milano,\\
Via Celoria 16, I-20133 Milano, Italy\\

\vspace*{0.5cm}

$^3$ INFN, Sezione di Milano,\\
Via Celoria 16,
I-20133 Milano, Italy\\
}

\end{center}

\vspace{1.5truecm}

\centerline{\bf ABSTRACT} \vspace{.5truecm} \noindent In this article we derive the full interacting effective actions for supersymmetric D-branes in arbitrary bosonic type II supergravity backgrounds. The actions are presented in terms of component fields up to second order in fermions. As one expects, the actions are built from the supercovariant derivative operator and the $\kappa$-symmetry projector. The results take a compact and elegant form exhibiting $\kappa$-symmetry, as well as supersymmetry in a background with Killing spinors. We give the explicit transformation rules for these symmetries in all cases, including the M2-brane. As an example, we analyze the N=2 super-worldvolume field theory defined by a test D4-brane in the supergravity background produced by a large number of  D0-branes.  This example displays rigid supersymmetry in a curved spacetime.

\newpage
\section{Introduction}

In the last decade, D-brane physics has become an important probe of  fundamental structures in string theory and M-theory. In particular, configurations of supersymmetric D-branes play a central role in the study of black hole entropy (e.g., 
\cite{SV}), the ADS/CFT correspondence (e.g., \cite{Juan}) and other gauge/gravity dualities, the Myers dielectric effect (e.g., 
\cite{myers}), matrix theory (e.g., \cite{matrix1,matrix2,matrix3}), and many other phenomena.
 Typically one concentrates on isolated D-branes where one can use the Born-Infeld
action, but see e.g.,  \cite{Bergshoeff:2001dc,Taylor:1999pr} for some work toward the non-abelian multi-brane generalization.

Most of the above studies have centered on the bosonic D-brane worldvolume fields.  While the full
actions (including fermions) for isolated supersymmetric D-branes have been described  \cite{Bergshoeff:1996tu,Cederwall:1996ri,Aganagic:1996pe} in terms of the superfield formalisms of type IIA and IIB supergravity, such descriptions leave the detailed fermion structure rather implicit.  Our purpose here (as for, e.g., \cite{tvr1,tvr2,MG}) is to make these features more explicit and to render the actions
in a form more useful for pedestrian calculations.

In a previous work \cite{mms1}, we obtained the explicit form of {\it partial} actions for supersymmetric D-branes in arbitrary bosonic backgrounds up to second order in fermions by means of the so-called ``normal coordinate expansion'' \cite{nc1}. Our previous actions were obtained in a particular limit of the worldvolume theory, in which interactions between the field combination ${\cal F}=b+F$ and the fermions could be ignored. Here, $b$ is the NS antisymmetric field and $F$ is the Yang-Mills field strength defined on the D-brane. In that work, the actions were found from that of the M2-brane \cite{gk1} by first performing a single dimensional reduction to the D2-brane and then computing the other brane actions by T-duality (adapted to the above approximation).
The results were claimed to be  $\kappa$-symmetric and to have supersymmetries as determined by
the background but, due to the different techniques involved, the
proofs were saved for the current paper.
Below we extend our previous work first by including the previously neglected interactions between ${\cal F}$ and the fermions, and second by including a detailed discussion of $\kappa$-symmetry and supersymmetry for the resulting D$p$-brane actions. We give an explicit expression for the $\kappa$-symmetric action of a single supersymmetric D$p$-brane in an arbitrary bosonic background, expanded to second order in the fermions, along with the corresponding $\kappa$-symmetry and supersymmetry transformation rules.

This article is organized as follows: In section 2 we derive the full interacting D$p$-brane actions. Starting once more from M2-brane, we perform a single dimensional reduction to obtain the D2-brane action. We then obtain the other D$p$-brane actions using T-duality following \cite{Bergshoeff:1995as,has,ms1}. The results appear in a form that is covariant under T-duality, but this form turns out not to be well adapted to the study of supersymmetry and $\kappa$-symmetry. 
We therefore begin anew in section three and derive the D2-brane action using a different reduction from the M2-brane, such that the fermions and in particular the $\kappa$-symmetry projector appear explicitly in the action.  This second approach relies on our previous work \cite{mms1} for technical input.  In the second part of this section, we obtain the other D$p$-brane actions using T-duality.   The argument also relies in part on the actions obtained in section 2.
After obtaining the new action in section 3, lengthy calculation shows that it agrees with the form given
in section 2.    While the derivations are not wholly independent, this agreement nevertheless provides a useful consistency check for our calculations.
  
In section 4 we present a detailed general discussion of $\kappa$-symmetry and supersymmetry for the above actions. Section five provides an explicit application to the case of a test D4-brane in the background produced by many D0-branes. Here we compute the form of the action and, fixing $\kappa$-symmetry, obtain the supersymmetry transformations of the world-volume field theory. We include three technical appendices:  a list of spinor definitions and conventions (appendix A), supergravity definitions and conventions (appendix B), and the detailed rules for the application of T-duality (appendix C).  Appendices A and B are essentially identical to those in our previous paper \cite{mms1} and are included here for completeness only.  Appendix C contains appendix C from \cite{mms1}, but also contains some further explanation of the calculations described in section four.

\section{A superfield-like form of the Dp-brane actions}\label{superactions}

In this section we begin with the M2-brane and, by means of a single dimensional reduction, obtain the D2-brane. Let us borrow the M2-brane action obtained in \cite{gk1} that results from an expansion in normal coordinates \cite{nc1,gz1} around an arbitrary bosonic background:
\bea
&&S_{M2}=S^{(0)}_{M2}+S^{(2)}_{M2}+O(y^{4}),\label{mt}\nn
&&S^{(0)}_{M2}=-T_{M2}\int{d^3\xi\sqrt{-\det(G)}}-
{T_{M2}\over 6}\int{d^3\xi\varepsilon^{ijk}A_{kji}},\label{m0}\nn
&&S^{(2)}_{M2}={iT_{M2}\over 2}\int d^3\xi\left\{\sqrt{-\det(G)}\left[ \bar{y}\Gamma^i\nabla_iy + \bar{y}T_i\Gamma^iy\right]+\right.\nonumber\nn
&&\;\;\;\;\;\;\;\;\;\;\;\;\;\;\;\;\;\;\;\;\;\;\;\;\;\;\;\;\;\;\;\;\;\;\;\;\
\left.-\;\hbox{${1\over 2}$}\varepsilon^{ijk}\left[\bar{y}\Gamma_{ij}\nabla_ky
+\bar{y}T_i\Gamma_{jk}y\right]\right\},
\label{M2}
\eea
where where $i=(0,1,2)$ are worldvolume indices along the brane, $T_{M2}=(4\pi^2 l_p^{3})^{-1}$, $l_p$ is the 11D Plank length, ($G,A$) are the pull-backs of the eleven dimensional metric and the gauge 3-form, $y$ is a real Majorana 32 component spinor,  $\Gamma_i$ are pull-backs of real gamma matrices, $\nabla_i$ is the pull-back of the usual spinor covariant derivative, $T_i$ is the pull-back of
\be
T_{\hat a}={1\over288}(\Gamma_{\hat a}^{\;\;\hat b \hat c \hat d \hat e}+8\delta^{\hat b}_{\hat a}\Gamma^{\hat c \hat d \hat e}) H_{\hat b \hat c \hat d \hat e}\;.
\ee
and $H=dA$. 

We begin by rewriting the membrane action (\ref{M2}) up to second
order in fermions in the form 
\bea
S^{M2}=-T_{M2}\int{d^3\xi\sqrt{-\det({\bf \hat G})}}- {T_{M2}\over
6}\int{d^3\xi\varepsilon^{ijk}{\bf \hat A}_{kji}},\label{sm}\nn
\eea 
where we have defined, 
\bea 
\label{11df}
{\bf \hat G}_{\hat m \hat n} &=& g_{\hat m \hat n}-i
{\bar y} \Gamma_{(\hat m}D_{\hat n)}y\,,\cr {\bf \hat A}_{\hat m \hat n \hat p}&=&
A_{\hat m \hat n \hat p}-\hbox{${3\over 2}$}i{\bar y}\Gamma_{[\hat m \hat n}D_{\hat p]}y. 
\eea
and ${\tilde D}_i=\nabla_i-{1\over288}(\Gamma_i^{\;\;bcde}-8\delta^b_i\Gamma^{cde})H_{bcde}$ is the pull-back of the supercovariant derivative of 11D supergravity (see appendix B for supergravity conventions).  Note that these fields are even in fermions and therefore, at least at the classical level, commute with all fields.

Our conventions for the reduction are to use hatted symbols for 11D indices and un-hatted symbols for 10D indices. We use $a,b,c\ldots=(0,1,\ldots,9)$ as tangent space indices and $m,n,o\ldots=(0,1,\ldots,9)$ as space-time indices. We also underline number indices corresponding to tangent space directions (i.e. $\Gamma^{\underline{0}}$), leaving space-time indices unadorned.

Thus the bosonic space-time coordinates $x^{\hat{m}}$ split into $(x^m,x^{10})$, where all the background fields are taken to be independent of $x^{10}$. As usual, one chooses a local frame for the reduction in which one has the vielbein
\bea
e_{\hat{m}}^{\;\;\hat{a}}=\left( \begin{array}{cc}
    e^{-\phi/3}e_m^{\;\;a} & -e^{2\phi/3}C_m  \\
    0 & e^{2\phi/3}
    \end{array}\right).
\eea
Here $e_m^{\;\;a}$ is the 10D vielbein in the string frame, $C^{(1)}=dx^mC_m$ is the RR 1-form potential, and $\phi$ is the dilaton. The 3-form potential of 11D supergravity $A=\hbox{${1\over 3!}$}dx^{\hat m}\wedge dx^{\hat n}\wedge dx^{\hat p}A_{\hat p \hat n \hat m}$ decomposes into the RR 3-form potential $C^{(3)}=\hbox{${1\over 3!}$}dx^m\wedge dx^n\wedge dx^pC_{pnm}$ and the NS two-form $b^{(2)}=\hbox{${1\over 2!}$}dx^m\wedge dx^nb_{nm}$ in the usual manner\footnote{ In this paper we always use the superspace convention for differential forms i.e. $w^{(p)}=\hbox{${1\over p!}$}dx^{m_1}\wedge\cdots \wedge dx^{m_p} w_{m_p\cdots m_1}$.},
\be
A_{mnp}=-C_{mnp} \;\;,\;\;A_{10\;mn}=b_{mn}\;.
\ee
In order that the supercoordinate transformations of 10D superspace maintain the canonical form, we also use the customary rescaling of fermions,
\be
y \longrightarrow e^{-{1\over6}\phi}y\; .
\ee
We introduce the following 10D fields,
\bea
\label{sfields} 
\Phi&=&\phi - \frac i2 \bar y \Delta y \cr
{\bf G}_{mn}&=& g_{mn}-i\bar y \Gamma_{(m}D_{n)}y \cr {\bf
B}_{mn}&=& b_{mn}-i\bar y \Gamma^{\ul{\varphi}}\Gamma_{[m}D_{n]}y
\cr {\bf C}_{mnp} &=& C_{mnp}+\frac i2 e^{-\phi}(3\bar y
\Gamma_{[mn}D_{p]}y-\bar y \Gamma_{mnp}\Delta)y\cr {\bf C}_m &=&
C_m +\bar y \Gamma^{\ul{\varphi}}(D_m-\Gamma_m\Delta) y
\eea
and the 10D operators\footnote{These operators appear in the supersymmetric transformations of the gravitino ($\delta \psi_m\sim D_m \epsilon$ ) and the dilatino ($\delta\lambda\sim \Delta \epsilon$) (see appendix B for supergravity conventions and definitions).}
\begin{eqnarray}
D_m &=& D^{(0)}_m+W_m \cr
\Delta &=& \Delta^{(1)}+\Delta^{(2)}\ ,
\end{eqnarray}
with
\begin{eqnarray}
D^{(0)}_m &=& \partial_m +\frac{1}{4} \omega_{mab}\Gamma^{ab}+\frac{1}{4\cdot 2!}H_{mab}\Gamma^{ab}\Gamma^{\ul{\varphi}} \cr
W_m &=& \frac18 e^\phi \left( \frac{1}{2!} {\bf F}^{(2)}_{ab}\Gamma^{ab}\Gamma_m\Gamma^{\ul{\varphi}}+
\frac{1}{4!}{\bf F}^{(4)}_{abcd}\Gamma^{abcd}\Gamma_m\right)\cr
\Delta^{(1)} &=& \frac12 \left( \Gamma^m \partial_m\phi +\frac{1}{2\cdot 3!}H_{abc}\Gamma^{abc}\Gamma^{\ul{\varphi}}\right)\cr
\Delta^{(2)}&=& \frac{1}{8} e^\phi \left( \frac{3}{2!} {\bf F}^{(2)}_{ab}\Gamma^{ab}\Gamma^{\ul{\varphi}}+
\frac{1}{4!} {\bf F}^{(4)}_{abcd}\Gamma^{abcd}\right)\ .
\end{eqnarray}
Except for the case of ${\bf C^{(3)}}$, these formulas are the result of reducing the 11-dimensional fields (\ref{11df}) in direct analogy with the reduction of purely bosonic fields.  Application of the standard reduction to  ${\bf A^{(3)}}$ would have produced the three-form
\begin{equation}
{\bf C}^{standard}_{mnp} = C_{mnp}+\frac i2 e^{-\phi}\bar y[3\Gamma_{[mn}D_{p]}
-\Gamma_{mnp}\Delta+6\Gamma^{\ul{\varphi}}\Gamma_{[m}C_nD_{p]}]y\, ,
\end{equation}
which contains an additional third term.

The reduction of the action (\ref{M2}) to obtain the D2-brane action is now carried out as it would be if one considered only the bosonic part of the membrane action. Since the form of our action is the same as that of the purely bosonic action, the reduction results in a D2-brane action with exactly the same form as for the bosonic D2-brane, but where each field contains a part that is a fermionic bilinear as described in (\ref{sfields}).  A small change then appears in replacing ${\bf C^{standard(3)}}$ by our ${\bf C^{(3)}}$.
The final form of the D2-brane action is then
\bea
\label{d2action4} 
S_{D2}= -T_{D2}\int d^3\xi
e^{-\Phi}\sqrt{-det({\bf G}+{\bf B} + F)}\;+ \cr
+\frac {T_{D2}}{6}\int
\epsilon^{ijk}{\bf C}_{kji} 
-\frac {T_{D2}}{2} \int
\epsilon^{ijk}{\bf C}_k{\cal F}_{ji}\ , 
\eea 
where $T_{D2}=(4\pi^{2}l_s^3g_s)^{-1}$ is the D2-brane tension, $l_s$ is the string length, $g_s$ is the string coupling and ${\cal F}_{ij}= b_{ij}+F_{ij}$. Note that (\ref{d2action4}) is indeed the action obtained by replacing the fields in the usual bosonic action by those of (\ref{11df}), except that in the final term ${\cal F}$ involves only the purely bosonic part of ${\bf B}$.  This is exactly the change induced by replacing ${\bf C^{standard(3)}}$ by ${\bf C^{(3)}}$.  It will prove useful in the discussion below.

\subsection{T-duality and the other Dp-branes}

Let us now derive the other D$p$-brane actions.
First, consider the Born-Infield part of (\ref{d2action4}), where the
new fields ${\bf G}_{mn}$, ${\bf B}_{mn}$ and $\Phi$ appear. The
corresponding fields for type IIB backgrounds in double spinor
notation (see appendixes \ref{conventions} and \ref{T-duality})
are 
\bea
\label{sfields2} 
\hat\Phi&=&\phi - \frac i2 \bar y \hat\Delta y \cr {\bf \hat G}_{mn}&=& g_{mn}-i\bar y
\Gamma_{(m}\hat D_{n)}y \cr {\bf \hat B}_{mn}&=& b_{mn}-i\bar y
{\hat \Gamma}^{\ul{\varphi}}\Gamma_{[m}\hat D_{n]}y \ . 
\eea
Using the T-duality rules of appendix \ref{T-duality}, it can be shown that these fields transform exactly as the pure bosonic ones \cite{ms1}.
More explicitly, performing a T-duality along the 9th direction we find that
\bea
&&\hat \Phi = \Phi -\frac 12 {\rm ln}\; {\bf G}_{99}\quad\quad\quad\quad\quad\qquad\quad\quad\quad\hspace{0.8cm}\hat{{\bf G}}_{ 99} = {1\over {\bf G}_{99}}
\nonumber\\
&&\hat{{\bf G}}_{ \tilde m \tilde n} = {\bf G}_{\tilde m \tilde n}
- { {\bf G}_{\tilde m 9} {\bf G}_{\tilde n 9}
- {\bf B}_{\tilde m 9} {\bf B}_{\tilde n 9}
\over {\bf G}_{ 99}}
\qquad\qquad
\hat{{\bf G}}_{\tilde m 9} ={ {\bf B}_{\tilde m 9}
\over {\bf G}_{99}}\\
&&\hat {\bf B}_{ \tilde m \tilde n}={\bf B}_{ \tilde m \tilde n}
-{{\bf B}_{\tilde m 9} {\bf G}_{\tilde n 9}-{\bf G}_{\tilde m 9}
{\bf B}_{\tilde n 9}\over {\bf G}_{99}} \quad\qquad\quad
\hat {\bf B}_{\tilde m 9} ={ {\bf G}_{\tilde m 9} \over {\bf G}_{
99}} \nonumber 
\eea 
where $\tilde m= 0,\ldots,8$. As a result, the
usual T-duality properties of the purely bosonic D$p$-brane
actions tell us that the Born-Infield part of the generic
D$p$-brane action is  the familiar-looking expression
\bea
\label{Born-Infeldaction} 
S_{Dp}^{Born-Infeld}&=& -T_{Dp}\int d^{p+1}\xi
e^{-\Phi}\sqrt{-det({\bf G}+{\bf B}+F)}\ , 
\eea 
where $T_{Dp}=2\pi[(2\pi l_s)^{p+1}g_s]^{-1}$ is the tension of the D$p$
brane.

To display the Ramond-Ramond part of the action it is natural to
define the new $p$-forms 
\bea 
\label{srrp}
{\bf C}^{(2p+1)}_{m_1\ldots
m_{2p+1}}= C_{m_1\ldots m_{2p+1}} 
+\frac i2 e^{-\phi}\big\{\bar y(\Gamma^{\ul{\varphi}})^{p+1}\big[ (2p+1) \Gamma_{[m_1\ldots m_{2p}}D_{m_{2p+1}]} \;-\cr  
-\Gamma_{m_1\ldots m_{2p+1}}\Delta \big]y\big\}\cr 
{\bf C}^{(2p)}_{m_1\ldots m_{2p}} = C_{m_1\ldots
m_{2p}}+\frac i2 e^{-\phi}\big\{\bar y(\hat\Gamma^{\ul{\varphi}})^{p}(i\sigma_2)\big[ (2p)\Gamma_{[m_1\ldots m_{2p-1}}\hat D_{m_{2p}]}-\quad\quad\cr 
-\Gamma_{m_1\ldots m_{2p}}\hat \Delta \big] y\big\}. 
\eea 
With a little patience, one may show that these fields satisfy the same T-duality rules as the purely bosonic Ramond-Ramond fields: 
\bea 
\tilde{ {\bf
C}}^{(n)}_{9\hat m_2\cdots \hat m_n}&=&{\bf C}^{(n-1)}_{\hat
m_2\cdots \hat m_n}-(n-1)g^{-1}_{99}g_{9[\hat m_2}{\bf
C}^{(n-1)}_{|9|\hat m_3\cdots \hat m_n]}\;, \nn \tilde{{\bf
C}}^{(n)}_{\hat m_1\cdots \hat m_n}&=&{\bf C}^{(n+1)}_{9\hat
  m_1\cdots \hat m_n}-nb_{9[\hat m_1}\tilde {\bf C}^{(n)}_{|9|\hat m_2\cdots \hat m_n]}\ .
\label{trr} 
\eea 
Interestingly, the form of (\ref{trr}) is not altered by the change from ${\bf C^{standard(3)}}$
to ${\bf C^{(3)}}$.

As a result, the Chern-Simons term of the D$p$-brane action is given by another familiar expression, 
\bea
S_{Dp}^{Chern-Simons}=T_{Dp}\int {\bf C}e^{-{\cal F}}\ , 
\label{sDp2} 
\eea
where ${\bf C}=\sum_n {\bf C}^{(n)}$ and we take the integral to
select the forms of the correct degree, $p+1$. 

\section{D$p$-brane actions and $\kappa$-symmetry projectors}\label{D2section}

In section 2, the D$p$-brane actions were obtained in a compact and elegant form.  However, this forms turn out not to be convenient for the study of $\kappa$-symmetry and supersymmetry.  In the current section, we organize the fermions in a different way so that the  $\kappa$-symmetry projector appears explicitly.   Guidance is best obtained by using known results for the M2-brane, so we return to 11-dimensions and start the reduction anew.  However, our calculation below will not be wholly independent of section 2, as we will borrow certain basic results concerning the structure of the final action.
It is important to note that the actions below are in fact identical to those presented in section \ref{superactions} and only differ in the way that certain terms are written. Nevertheless, these new forms are better adapted for the study of supersymmetry and $\kappa$-symmetry as will be seen in section \ref{symmetries}. 

As in section \ref{superactions} we begin with the M2-brane action and, by means of a single dimensional reduction, obtain the D2-brane. Following \cite{mms1}, we introduce the operator
\be\label{gammaM2}
\Gamma_{M2}=\hbox{${1\over 3!\sqrt{-G}}$}\epsilon^{ijk}\Gamma_{ijk}\ ,
\ee
so that the action (\ref{M2}) can be rewritten in the more compact and suggestive form
\bea
S_{M2}=-T_{M2}\int{d^3\xi\sqrt{-\det(G)}}-
{T_{M2}\over 6}\int{d^3\xi\varepsilon^{ijk}A_{kji}}+\cr
+\frac{iT_{M2}}{2}\int d^3\xi\sqrt{-G}\bar{y}(1-\Gamma_{M2})\Gamma^i\tilde{D}_iy\ .
\label{m2}
\eea
This action is by construction $\kappa$-symmetric up to higher order fermion terms and, as a result, is invariant under supersymmetries corresponding to any bulk Killing spinor. Again, the explicit transformation rules will be displayed in section \ref{symmetries}. Note that $\Gamma_{M2}$ squared is the identity operator.

We now perform a dimensional reduction to obtain the corresponding action for the type IIA D2-brane in 10D (up to second order in $y$). 
Here the results are more complicated and less familiar in appearance than those of section 2, so we explain the process in correspondingly more detail.

Using the Kaluza-Klein framework of the previous section, we introduce the useful world-volume one-form
\be
p_i=\partial_ix^{10}-\partial_ix^mC_m\ .
\ee
The pull-back of the bosonic 11D metric $G$ may now be written
\be
G_{ij}=e^{-2\phi/3}g_{ij}+e^{4\phi/3}p_ip_j\;,
\ee
where $g$ is the 10D metric.
In a background of our Kaluza-Klein form, the M2-brane action becomes
\bea\label{d2action1}
&&S_{M2}=T_{D2}\Big\{-\int d^3\xi \sqrt{-g(1+e^{2\phi}p^2)}(1-\frac i2 \bar y \Gamma^i D_i y)\;\;+\cr
&&+\frac i2\int \frac{e^{-\phi}\sqrt{-g}}{\sqrt{1+e^{2\phi}p^2}}\big[ e^\phi p^i\bar y \Gamma^{\ul{\varphi}}(D_i -\Gamma_i\Delta)y-\bar y \Delta y -e^{2\phi}p^ip^j \bar y \Gamma_i D_j y\big]\cr
&&-\frac 12 \int \epsilon^{ijk}p_k(b_{ji}-i\bar y \Gamma^{\ul{\varphi}}\Gamma_j D_i y)-\frac i2 \int d^3\xi\sqrt{-g}\bar y \Gamma_{D2}(\Gamma^i D_i -\Delta)y+\cr
&&+\frac 16\int \epsilon^{ijk}C_{kji}-\frac 12 \int \epsilon^{ijk}C_kb_{ji}\Big\}\ ,
\eea
where $\Gamma_{D2}=\hbox{${1\over 3!\sqrt{-g}}$}\epsilon^{ijk}\Gamma_{ijk}$.

Let us recall that to obtain the D2-brane action with its characteristic Yang-Mills field strength $F=dA$, we must replace the scalar $p_i$ by its world-volume dual, the gauge 1-form potential $A$ \cite{Bergshoeff:1996tu}. To do so, we promote $p_i$ to an independent variable and compensate by adding a Lagrange multiplier term $\hbox{${1\over2}$}\epsilon^{ijk}(p_i+C_i)F_{jk}$ to the M2-brane action (\ref{d2action1}). We then solve for the $p_i$'s using the equations of motion obtained by varying these same $p_i$'s.  
Finally, the result is inserted back into the original action.
This procedure produces the D2-brane action 
\bea
\label{d2action2}
S_{D2}&=& T_{D2}\Big\{ -\int d^3\xi e^{-\phi}\sqrt{-det(g+{\cal F})}+\frac 16\int \epsilon^{ijk}C_{kji} -\frac 12 \int
\epsilon^{ijk}C_k{\cal F}_{ji}\,+\cr 
&& +\frac i2 \int d^3\xi e^{-\phi}\sqrt{-det(g+{\cal F})}\bar y(1-\frac{\Gamma_{D2}}{\sqrt{1+{\cal F}}})(\Gamma^i D_i
-\Delta)y\,+\cr 
&& +\frac i2\int d^3\xi e^{-\phi}\frac{\sqrt{-det(g)}}{\sqrt{1+{\cal F}}}{\cal F}^{ij}\bar y \Gamma^{\ul{\varphi}}\Gamma_i D_j y +\cr
&&- \frac i4 \int d^3\xi e^{-\phi}\epsilon^{ijk}{\cal F}_{kj}\bar y\Gamma^{\ul{\varphi}}(D_i-\Gamma_i\Delta) y\,+\cr && 
+\frac i2\int d^3\xi e^{-\phi}\frac{\sqrt{-det(g)}}{\sqrt{1+{\cal F}}}{\cal F}^{ik}{\cal F}_k{}^j\bar y \Gamma_i D_j y \Big\}\ , 
\eea 
where by $1+{\cal F}$ we mean $det(\delta^i_j+{\cal F}^i{}_j)$.

\subsection{The other D-brane actions}\label{actions}

We now derive the remaining D-brane actions using the well
established T-duality properties of Hassan's supergravity
formalism \cite{has}. As a bonus, the expressions that we obtain
throughout this procedure come in a form for which
$\kappa$-symmetry and supersymmetry are almost evident (both
symmetries will be studied in detail in section \ref{symmetries}).

We begin by introducing a new chirality operator,
\bea
\label{gammaD2tilde}
\tilde\Gamma_{D2}&=&\frac{1}{\sqrt{1+{\cal
F}}}(\Gamma_{D2}+\frac{{\cal
F}^{ij}}{2\sqrt{-g}}\epsilon_{ij}{}^k\Gamma_k\Gphi)=\cr
&=&\frac{1}{\sqrt{1+{\cal F}}}(1-\frac12 {\cal
F}^{ij}\Gamma_{ij}\Gphi)\Gamma_{D2}\ , 
\eea 
that allows us to rewrite the D2-brane action
(\ref{d2action2}), in the more convenient form
\bea
\label{d2fermion}
S_{D2}^{(2)}=\frac {iT_{D2}}{2}\int d^3\xi
e^{-\phi}\sqrt{-(g+{\cal F})}\bar y
(1-\tilde\Gamma_{D2})(\Gamma^iD_i -\Delta\, +\cr
-\frac{\epsilon^{ijk}}{2\sqrt{-(g+{\cal F})}}{\cal F}_{ij}\Gphi
\Gamma_k{}^rD_r)y\ ,
\eea 
where the superscript (2) on the left hand side indicates that we have displayed that part of the action which is second order in fermions.

The equivalence between the two forms of $S_{D2}^{(2)}$
(\ref{d2action2},\ref{d2fermion}) can be seen as follows: first is trivial to check that in (\ref{d2fermion}) the first two terms of the projector ${1\over 2}(1-\tilde\Gamma_{D2})$ multiplied by  the term $\Gamma^iD_i -\Delta$ corresponds to the second line in the action (\ref{d2action2}).
Next,  the second and third term of the projector times the term $-\frac{\epsilon^{ijk}}{2\sqrt{-(g+{\cal F})}}{\cal F}_{ij}\Gphi \Gamma_k{}^rD_r$ correspond to first term in the third line of (\ref{d2action2}) and the term in the fourth line of (\ref{d2action2}). All that then remains is to compute the product of the first term of the projector times $-\frac{\epsilon^{ijk}}{2\sqrt{-(g+{\cal F})}}{\cal F}_{ij}\Gphi \Gamma_k{}^rD_r$ plus the third term of the projector times $\Gamma^iD_i -\Delta$, but this gives exactly the missing term
 $\epsilon^{ijk}{\cal F}_{kj}\bar y \Gamma^{\ul{\varphi}}(D_i-\Gamma_i\Delta)$ in (\ref{d2action2}), completing the argument.

We now wish to T-dualize $S_{D2}^{(2)}$.
Let us begin by writing (\ref{d2fermion}) as 
\bea
\label{part1} 
\frac {iT_{D2}}{2}\int d^3\xi e^{-\phi}\sqrt{-(g+{\cal F})}\bar y (1-\tilde\Gamma_{D2})(\Gamma^iD_i - \Delta + O({\cal F}))y \,. 
\eea 
The T-duality rules are given in appendix C. After some calculation it can be seen that under T-duality $\tilde\Gamma_{D2}$ transforms into the operator $\tilde\Gamma_{Dp}$ given by
\bea
\label{Ggeneral}
\tilde\Gamma_{D(2n)}=\hbox{$\frac{1}{\sqrt{-(g+{\cal F})}}$}\sum_{q+r=n}\hbox{$\frac{\epsilon^{i_1\ldots i_{2q}j_1\ldots j_{2r+1}}}{q!2^q(2r+1)!}$}{\cal F}_{i_1i_2}\cdots{\cal F}_{i_{2q-1}i_{2q}}\Gamma_{j_1\ldots j_{2r+1}}(\Gphi)^{r+1}\ ,\cr
\tilde\Gamma_{D(2n+1)}=\hbox{$\frac{-i\sigma_2}{\sqrt{-(g+{\cal F})}}$}\sum_{q+r=n+1}\hbox{$\frac{\epsilon^{i_1\ldots i_{2q}j_1\ldots j_{2r}}}{q!2^q(2r)!}$}{\cal F}_{i_1i_2}\cdots{\cal F}_{i_{2q-1}i_{2q}}\Gamma_{j_1\ldots j_{2r}}(\hGphi)^{r}.
\eea
These operators coincide with the supersymmetric zero order term, in a fermionic expansion, of the superoperator found in \cite{Bergshoeff:1996tu}. In addition, these $\tilde\Gamma_{Dp}$ are the operators that will appear in the D$p$-brane $\kappa$-symmetry transformations presented at the end of section \ref{symmetries}. 

On the other hand, in \cite{mms1} we showed that the term $(\Gamma^iD_i -\Delta)$ is invariant up to terms containing ${\cal F}$. Thus, after T-duality $S^{(2)}_{D2}$ will again be of the form $(\Gamma^iD_i -\Delta+ O({\cal F}))$.  The $O({\cal F})$ terms appearing in the D$p$-brane actions can be obtained through lengthy calculation and involve the operators
\bea
L_{2n+1}&&=\sum_{q\geq 1,q+r=n+1}\hbox{$\frac{\epsilon^{i_1\ldots i_{2q}j_1\ldots j_{2r}}(-i\sigma_2)(\hGphi)^{r}}{q!2^q(2r)!\sqrt{-(g+{\cal F})}}{\cal F}_{i_1i_2}\cdots{\cal F}_{i_{2q-1}i_{2q}}\Gamma_{j_1\ldots j_{2r}}{}^k \hat D_k$} \ ,\cr
L_{2n}=&&\sum_{q\geq 1,q+r=n}\hbox{$\frac{\epsilon^{i_1\ldots i_{2q}j_1\ldots
    j_{2r+1}}(-\Gphi)^{r+1}}{q!2^q(2r+1)!\sqrt{-(g+{\cal F})}}$}\hbox{${\cal
  F}_{i_1i_2}\cdots{\cal F}_{i_{2q-1}i_{2q}}\Gamma_{j_1\ldots j_{2r+1}}{}^k D_k$}
\eea
for D(2$n$+1)- and D(2$n$)-branes respectively.  Here we simply quote the final result, but more information can be found in appendix C.1.  In particular, C.1 explains how the calculation can be somewhat simplified by taking one basic cue from section 2.
The result is
\begin{equation}
\label{Dpt}
S^{(2)}_{Dp}=\frac
{iT_{Dp}}{2}\int d^{p+1}\xi e^{-\phi}\sqrt{-(g+{\cal
    F})}\bar y(1-\tilde\Gamma_{D_p})(\Gamma^iD_i-\Delta+L_p)y\ .
\end{equation}
where the hat on the operators $D,\Delta$ is understood for the type IIB branes. Therefore, the full D$p$-brane action is: 
\bea
S_{Dp}&=&S^{(0)}_{Dp}+S^{(2)}_{Dp}+O(y^{4}),\nn
S_{Dp}^{(0)}&=& -T_{Dp}\int d^{p+1}\xi e^{-\phi}\sqrt{-(g+{\cal
F})}+T_{Dp}\int C\; e^{-{\cal F}}\,, \cr 
S^{(2)}_{Dp}&=&\frac
{iT_{Dp}}{2}\int d^{p+1}\xi e^{-\phi}\sqrt{-(g+{\cal
    F})}\bar y(1-\tilde\Gamma_{D_p})(\Gamma^iD_i-\Delta+L_p)y .
\label{dpactions2}
\eea

These actions are indeed equivalent to the ones obtained in section
\ref{actions}, though some calculation is required to confirm this.  In particular, while the equivalence is straightforward for the Chern-Simons term (\ref{sDp2}), the terms coming from the Born-Infield action are more subtle for $p\geq 3$ and one must use some nontrivial Fierz-like identities. For example, for the D$3$-brane, after having expanded the action (\ref{Born-Infeldaction}) one requires the identities 
\bea 
{\cal F}_{ik}{\cal F}^{kr}{\cal
F}_{rj}&=& -\frac 12 {\cal F}^{rs}{\cal F}_{rs}{\cal
F}_{ij}+\frac{\epsilon^{lstw}}{16 (det g)}{\cal F}_{ls}{\cal F}_{tw}{\cal
F}^{kr}\epsilon_{krij} \ \ {\rm and} \cr {\cal F}^{ik}{\cal
F}_{kr}{\cal F}^{rs}{\cal F}_{sj}&=& -\frac 12 {\cal F}^{rs}{\cal
F}_{rs}{\cal F}^{ik}{\cal
F}_{kj}-\frac{1}{64(det g)}(\epsilon^{lstw}{\cal F}_{ls}{\cal
F}_{tw})^2\delta^i_j\ .
\eea 
However, in the end one does indeed show that the two actions agree.

\section{$\kappa$-symmetry and Supersymmetry}\label{symmetries}

In this section we obtain the $\kappa$-symmetry and supersymmetry
transformations of the world-volume fields. We begin by
considering such transformations in the framework of
the normal coordinate expansion. In what follows we use the
analysis developed in \cite{gz1}, modified to accommodate our
particular needs\footnote{The analysis of \cite{gz1} was carried out in the context of heterotic string theory.}.

In the superspace formalism, the supercoordinates $z^M$ decompose
into bosonic coordinates $x^m$ and fermionic coordinates
$\theta^\mu$. Here we also introduce a similar decomposition for
tangent space vectors $y^A$, with $A=(a,\alpha)$: \bea
&&z^M=(x^m,\theta^\mu), \nn &&y^A=(y^a,y^\alpha). \eea The normal
coordinate expansion is a method based on the definition of normal
coordinates in a neighborhood of a given point $z^M$ of
superspace. The idea is to parameterize the neighboring points by
the tangent vectors along the geodesics joining these points to
the origin. Denoting the coordinates at neighboring points by
$Z^M$ and the tangent vectors by $y^A$, we have \be
Z^M=z^M+\Sigma^M(y)\;, \ee where the explicit form of
$\Sigma^M(y)$ is found iteratively by solving the geodesic
equation. Tensors at the point $Z^M$ may be compared with those at
$z^M$ by parallel transport.  In this sense, the change in a
general tensor under an infinitesimal displacement $y^A$ is \be
\delta T = y^A\nabla_A T\; . \label{exp1} \ee Finite transport is
obtained by iteration. In this way we may consider the
corresponding expansion  in the operator $\delta$ for any tensor
in superspace. For example, consider the vielbein $E_M^A$ \be
E_M^{\;\;\;A}(Z)=E_M^{\;\;\;A}(z)+ \delta E_M^{\;\;\;A}(z)+
{1\over2}\delta^2E_M^{\;\;\;A}(z)+\ldots \ee In particular, in 11D
supergravity one finds the following fundamental relations by
means of which one can expand any tensor iteratively to any order
(see for example \cite{gk1}): \bea &&\delta E^{\hat a} =0, \nn
&&\delta E^\alpha = Dy^\alpha +y^\beta e^{\hat b} T_{\hat b
\beta}^{\;\;\;\alpha}\;, \nn &&\delta^2 E^{\hat a}
=-i(y^\alpha\Gamma^{\hat a}_{\;\;\alpha\beta}Dy^\beta+y^\alpha
y^\beta T_{\;\beta}^{\;\;\;\gamma}\Gamma^{\hat
a}_{\;\;\gamma\alpha})\;, \nn &&\delta^2 E^\alpha = 0\;.
\label{exp2} \eea In the above, $T$ is the torsion and $R$ the
Riemann tensor (see appendix B for supergravity conventions). In
our case, we are interested in an expansion up to second order
around a bosonic background. Therefore, we have set $z^M=(x^m,0)$
and $y^A=(0,y^\alpha)$ in equation (\ref{exp2}).

As can be seen from the short introduction above, the normal
coordinate expansion defines the coordinates of an arbitrary point
in superspace $Z^M=(x^m,y^\alpha)$ by, first, choosing an origin
in superspace of the particular form $z^M=(x^m,0)$ and, second,
obtaining $y^\alpha$ by way of the geodesic joining the two
points. The result is that transformations defined in terms of
superfields can be defined  as a transformation of $Z^M$, of $z^M$, or of both. In our
case, we wish to preserve the purely bosonic form of $z^m$,  and
the purely fermionic separation of the two points along the
geodesic, together with the Wess-Zumino gauge for the
supervielbein, i.e. \bea E(z)_{M}^{\;\;A}=\left(
\begin{array}{cc}
    e(x)_{\hat m}^{\;\;\hat a} & \psi(x)_{\hat m}^{\;\;\alpha}  \\
    0 & \delta_\mu^{\;\;\alpha}
    \end{array}\right).
\eea These constraints single out a particular transformation
below which acts on both $z^M$ and $Z^M$.

\subsection{M2-brane}

Let us begin with the transformation laws of the supersymmetric
M2-brane in 11D superspace. In this case, $\kappa$-symmetry
\cite{Bergshoeff:1987cm} is given by 
\be 
\delta Z^ME_M^{\
a}=0\;\;\; {\rm and}\;\;\;\delta Z^ME_M^{\
\alpha}=(1+\Gamma)\kappa. 
\ee 
where $\Gamma$ has the same form as $\Gamma_{M2}$ but is constructed with the pull-back of full supervielbein $E$.
After using the 11D supergravity
constraints, the transformation rules that preserve the above choices of gauge can be computed following the standard procedure of \cite{gz1}. Here we give the final form in a general
bosonic background\footnote{Since we are discussing symmetries of the worldvolume theory, the supergravity background fields $L$ are fixed and therefore transform as scalar fields, i.e. $\delta L
=\delta x^{\hat m}\partial_{\hat m} L(x)$. Therefore, in the rest of the paper this rule will be omitted.}, 
\bea
\label{km2} \delta_{\kappa}y&=&(1+\Gamma_{M2})\kappa+O(y^2)\ , \cr
\delta_{\kappa}x^{\hat m}&=&\frac i2 \bar y \Gamma^{\hat m}
(1+\Gamma_{M2})\kappa+O(y^3)\ .
\eea 
The general superspace transformation is given by
\be 
\delta Z^M=\epsilon^M(Z). 
\ee 

We are most interested in the
case where the bosonic background breaks only a fraction of the
total supersymmetries; i.e. when some spinor $\varepsilon$
satisfies the killing spinor equation $\tilde D_{\hat m}\varepsilon =0$.
Superspace transformations corresponding to $\varepsilon$  leave
the background invariant and induce supersymmetry transformations
that leave the M2-brane action similarly invariant. Following the
standard procedure of \cite{gz1} we obtain the explicit rules,
\bea 
\label{sm2} 
\delta_{\varepsilon}y&=&\varepsilon+O(y^2)\ , \cr
\delta_{\varepsilon}x^{\hat m}&=&-\frac i2 \bar y \Gamma^{\hat m}
\varepsilon +O(y^3). 
\eea 
The normal coordinate expansion by construction guaranties that the
 transformations (\ref{km2},\ref{sm2}) are indeed invariances
of the worldvolume theory, but we have also checked explicitly
that the action (\ref{d2action2}) is unchanged by (\ref{sm2}).

\subsection{D$p$-branes}

Let us now study supersymmetry and $\kappa$-symmetry for the D2
brane. Considering first supersymmetry, from direct reduction of
(\ref{sm2}) one obtains the transformation rules for the spinor
$y$ and the embedding fields $x^m$: 
\bea 
\label{sd2}
\delta_{\varepsilon}y&=&\varepsilon+O(y^2)\ , \cr
\delta_{\varepsilon}x^m&=&-\frac i2 \bar y \Gamma^m \varepsilon
+O(y^3)\ . 
\eea 
To obtain the transformation rules for the gauge
field $A_i$ we use  a trick introduced in
\cite{Bergshoeff:1996tu}. One starts from the observation that the
action (\ref{d2action1}) is supersymmetric thanks to the property
$d(p^{(1)}+C^{(1)})=0$. Once $p_i$ is considered as an independent
variable, the supersymmetry variation of the action
(\ref{d2action1}) must have $\epsilon^{ijk}\partial_j (p_k +C_k)$
as a factor and 
the supersymmetry of $S^{D2}$ is ensured by an appropriate
variation of the gauge field $A_i$. In this context, the relevant
terms in (\ref{d2action1}) are those coming from the Chern-Simons
term and, with a little inspection, one realizes that the
transformation rules for the gauge field are 
\bea
\label{sA}
\delta_{\varepsilon}A_i=\frac i2 \bar y
\Gamma^{\ul{\varphi}}\Gamma_i \varepsilon -\frac i2 b_{im}\bar y
\Gamma^m\varepsilon +O(y^3)\ . 
\eea

We obtain the transformation rules for the general D$p$ brane by performing T-duality.  It turns out to be simplest not to calculate covariantly but, instead, to first
impose static gauge on our D2-brane.  
After performing the T-duality, it is not difficult to write down the unique covariant expression associated with our static gauge result. In static gauge one imposes the
condition $x^i(\xi)=\xi^i$ for $i=0,1,2$, so that we have to
compensate the supersymmetry transformation of the fields with a
worldvolume diffeomorphism, $\delta\xi^i=-\frac i2 \bar y \Gamma^i
\epsilon$. Hence, up to a gauge transformation of $A_i$, the supersymmetries
transformations in the static gauge become 
\bea
\delta_{\varepsilon}y&=&\varepsilon+O(y^2)\ , \cr
\delta_{\varepsilon}x^{\tilde m}&=&-\frac i2 \bar y \Gamma^{\tilde
m} \varepsilon+\frac i2 (\partial_i x^{\tilde m})\bar y \Gamma^i
\varepsilon +O(y^3)\cr \delta_{\varepsilon}A_i&=&\frac i2 \bar y
\Gamma^{\ul{\varphi}}\Gamma_i \varepsilon -\frac i2 b_{im}\bar y
\Gamma^m\varepsilon-\frac i2 F_{ij}\bar y \Gamma^j \varepsilon
+O(y^3)\ , 
\eea 
where $\tilde m=3,\dots,9$. At this point, one can
perform T-duality, obtaining the supersymmetry rules for the
D3-brane in the static gauge 
\bea
\delta_{\varepsilon}y&=&\varepsilon+O(y^2)\ , \cr
\delta_{\varepsilon}x^{\hat m}&=&-\frac i2 \bar y \Gamma^{\hat m}
\varepsilon+\frac i2 (\partial_I x^{\tilde m})\bar y \Gamma^I
\varepsilon +O(y^3)\cr \delta_{\varepsilon}A_I&=&\frac i2 \bar y
\hat \Gamma^{\ul{\varphi}}\Gamma_I \varepsilon -\frac i2 b_{Im}\bar y
\Gamma^m\varepsilon-\frac i2 F_{IJ}\bar y \Gamma^J \varepsilon
+O(y^3)\ , 
\eea 
where now $I,J=0,\ldots,3$, $\hat m=4,\ldots,9$
and we are using the double spinor convention for the chiral IIB
case summarized in appendix \ref{conventions}. These
transformations are obtained as specialized to the static gauge
for the D3-brane, but it is easy to write them in the covariant
form 
\bea
\label{generalsusy}
\delta_{\varepsilon}y&=&\varepsilon+O(y^2)\ , \cr
\delta_{\varepsilon}x^{m}&=&-\frac i2 \bar y \Gamma^{m}
\varepsilon+O(y^3)\cr \delta_{\varepsilon}A_I&=&\frac i2 \bar y
\hat \Gamma^{\ul{\varphi}}\Gamma_I \varepsilon -\frac i2 b_{Im}\bar y
\Gamma^m\varepsilon +O(y^3)\ .
\eea 
Note that the index $m$ runs
over the entire set of 10D directions and there is no need to introduce
any extra diffeomorphism transformation since we have abandoned the
static gauge.
Iterating this procedure, we find that the supersymmetry transformations for any D$p$-brane are always of the form (\ref{generalsusy}), where for $p$ odd we use the double spinor convention and $\hGphi$ instead of $\Gphi$.

To study $\kappa$-symmetry, one proceeds identically . Reducing
the M2 operator $\Gamma_{M2}$ one obtains a third operator for the
D2 brane 
\bea 
\hat\Gamma_{D2}= \Gamma_{M2}=\sqrt{1+\frac 12 {\cal
F}^{ij}{\cal F}_{ij}} \Gamma_{D2}-\frac 12 {\cal
F}^{ij}\Gamma_{ij}\Gamma^{\ul{\varphi}}\ . 
\eea 
Then, the $\kappa$-symmetry transformations take the form: 
\bea 
\label{kd2}
\delta_{\kappa}y&=&(1+\hat\Gamma_{D2})\kappa +O(y^3)\ , \cr
\delta_{\kappa}x^m&=&\frac i2 \bar y \Gamma^m(1+\hat\Gamma_{D2})
\kappa +O(y^3)\cr 
\delta_{\kappa}A_i &=&-\frac i2 \bar y
\Gamma^{\ul{\varphi}}\Gamma_i(1+\hat\Gamma_{D2}) \kappa +\frac i2
b_{im}\bar y \Gamma^m(1+\hat\Gamma_{D2})\kappa+O(y^3)\ . 
\eea 
It should be emphasized that the operator $\hat\Gamma_{D2}$ is
different from the operator $\tilde\Gamma_{D2}$ which naturally
appears in the D2 brane action of the previous section and which
is so well suited to  T-duality. In contrast,
$\hat\Gamma_{D2}$ turns out not to transform nicely under T-duality. But
the two operators are related by 
\bea 
1+\hat\Gamma_{D2}=\sqrt{1+{\cal F}}\Gamma_{D2}(1+\tilde\Gamma_{D2})\
. 
\eea 
For this reason it is convenient to rewrite the
$\kappa$-symmetry transformation rules in the form 
\bea
\label{kd2bis} \delta_{\kappa}\bar
y&=&\bar\kappa(1+\hat\Gamma_{D2}) +O(y^3)\ , \cr
\delta_{\kappa}x^m&=&-\frac i2 \bar\kappa
(1+\hat\Gamma_{D2})\Gamma^m y +O(y^3)\cr 
\delta_{\kappa}A_i &=&\frac i2 \bar\kappa
(1+\hat\Gamma_{D2})\Gamma^{\ul{\varphi}}\Gamma_i y -\frac i2
b_{im}\bar \kappa (1+\hat\Gamma_{D2})\Gamma^m y + O(y^3)\ . 
\eea 
As we have just said, these rules do not transform nicely under
T-duality.  However,  since $\bar\kappa$ is an arbitrary spinor we
can redefine it by absorbing the operator $\sqrt{1+{\cal
F}}\Gamma_{D2}$ acting on it from the right.  In terms of this new
spinor, the the $\kappa$-symmetry transformations take the form
\bea 
\label{kd2tris} \delta_{\kappa}\bar
y&=&\bar\kappa(1+\tilde\Gamma_{D2}) +O(y^3)\ , \cr
\delta_{\kappa}x^m&=&-\frac i2 \bar\kappa
(1+\tilde\Gamma_{D2})\Gamma^m y +O(y^3)\cr 
\delta_{\kappa}A_i &=&\frac i2 \bar\kappa
(1+\tilde\Gamma_{D2})\Gamma^{\ul{\varphi}}\Gamma_i y -\frac i2
b_{im}\bar \kappa (1+\tilde\Gamma_{D2})\Gamma^m y +O(y^3)\ . 
\eea 
At this point, we can proceed as for supersymmetry: fix the static
gauge, perform the T-duality, and relax the static gauge to obtain
a covariant form. The result for the general D$p$-brane is 
\bea
\label{kDp} 
\delta_{\kappa}\bar
y&=&\bar\kappa(1+\tilde\Gamma_{Dp}) +O(y^3)\ , \cr
\delta_{\kappa}x^m&=&-\frac i2 \bar\kappa
(1+\tilde\Gamma_{Dp})\Gamma^m y +O(y^3)\cr \delta_{\kappa}A_i
&=&\frac i2 \bar\kappa
(1+\tilde\Gamma_{Dp})\Gamma^{\ul{\varphi}}\Gamma_i y -\frac i2
b_{im}\bar \kappa (1+\tilde\Gamma_{Dp})\Gamma^m y+O(y^3)\ , 
\eea
where the hat over $\Gphi$ for $p$ odd is understood.

\section{Example: the D0/D4 case}

In this section we show how the above actions and techniques can be applied to study interesting brane physics. In particular, consider the case of a single test D4-brane living in the space-time generated by a large number of D0-branes. The number of surviving supercharges is 8, $1/4$ of the maximal 32.

The D0-brane background is given by
\bea
&ds^2=e^ae^b\eta_{ab}\;\;\;,\;\;\;e^\phi=H^{-3/4}\;\;\;,\;\;C_0=H^{-1}-1\ ,&\cr
&e^{\underline 0}=H^{-1/4}dt\;\;\;,\;\;\;
e^{\hat a}=H^{1/4}\delta^{\hat a}_{\hat m}dx^{\hat m}\ .&
\eea  
where ($\hat m,\hat n, ...$) are curved-space indices, and ($\hat a,\hat b, ...$) are tangent space indices, both running from 1 to 9.
In addition,  $H$ is a harmonic function on $\mathbb{R}^9$ and $e^a$ are the vielbeins. Note that this background has $16$ surviving supersymmetries, given by the Killing spinor $\varepsilon=H^{-1/8}\varepsilon_0$ with $\varepsilon_0$ a constant 32 component Majorana spinor satisfying the equation $(1+\Gamma_{\underline 0}\Gamma^{\underline \varphi})\varepsilon_0=0$.

To write the D4-brane action in the above background, we need to compute the explicit form of the operators $D_m,\Delta$
\bea
D_m&=&\nabla_m+{1\over 16}e^\phi F_{mn}\Gamma^{mn}\Gamma^{\underline \varphi}\ ,\cr
\Delta&=&{1\over 2}\Gamma^m\partial_m \phi+ {3\over 16}e^\phi F_{mn}\Gamma^{mn}\Gamma^{\underline \varphi} \ .
\eea 
Writing the above in terms of the harmonic funtion $H$, one finds
\bea
D_t&=&\partial_t+{1\over8}H^{-3/2}\partial_{\hat n}H\delta^{\hat n}_{\hat a}\Gamma^{\hat a}\Gamma_{\underline 0}(1+\Gamma_{\underline 0}\Gamma^{\underline \varphi}) \ , \cr
D_{\hat m}&=&\partial_{\hat m}+{1\over8}\partial_{\hat n}\ln H 
-{1\over8}\partial_{\hat n}\ln H\delta^{\hat n}_{\hat b}\delta^{\hat a}_{\hat m}\Gamma^{\hat b}\Gamma_{\hat a}(1+\Gamma_{\underline 0}\Gamma^{\underline \varphi})\ , \cr
\Delta&=&{3\over 8}H^{-5/4}\partial_{\hat m}\delta^{\hat m}_{\hat a}(1+\Gamma_{\underline 0}\Gamma^{\underline \varphi})\ .
\eea

Since we are interested in the worldvolume field theory defined on the D4-brane, we choose the static gauge such that the worldvolume coordinates $\xi^i$ are identified with the background coordinates as follows: $\xi^0=t$, $\xi^I=x^I$, where $i=(0,I)$ and $I$ runs from 1 to 4. We also rescale the fields,
\be
A_i\rightarrow \lambda A_i\;\;,\;\;
x^{\dot m} \rightarrow \lambda \Phi^{\dot m}\;\;,\;\;y \rightarrow \lambda y\ ,
\ee
and take the limit $\lambda\rightarrow 0$, to obtain more standard normalizations and a non-interactive curved spacetime field theory, where the supersymetry will be linearly realized.  
Here $\dot m$ runs from 5 to 9, representing directions transverse to the D4-brane, and $\lambda= 2\pi\alpha'$.

After expanding the D4-brane action (either (\ref{Born-Infeldaction}) plus (\ref{sDp2}) or (\ref{dpactions2})) in this background one finds
\bea
\label{D4inD0}
&&S_{D4}=S^{(0)}_{D4}+S^{(2)}_{D4}+O(y^{4},\lambda^1),\nn
&&S_{D4}^{(0)}= -T_{D4}\int d^5\xi 
\ \ -\hbox{${1\over g^2_{YM}}$}\int d^5\xi\left\{ \hbox{${1\over 4}$}F^{ij}F_{ij} + \hbox{${1\over2}$}\partial^i\Phi^{\dot m}\partial_i\Phi^{\dot n}g_{\dot m \dot n} \ +\right . \cr
&&\hspace{7.5truecm}\left . - \hbox{${1\over 8}$}\Theta\epsilon^{0IJKL}F_{IJ}F_{KL}\right\}\ , \cr 
&&S^{(2)}_{D4}=\hbox{$\frac{i}{2g^2_{YM}}$}\int d^5\xi \left\{\bar y(1-\Gamma)\left[\Gamma^i\partial_i-\hbox{${1\over 8}$}\partial_I\ln H \Gamma^I(1+2\Gamma_{\underline0}\Gamma^{\underline \varphi})\right]y\right\}\ .
\eea
Here for simplicity we have kept only the terms of lowest nontrivial order in $\lambda$.
In addition we have introduced
$g^2_{YM}=(2\pi)^2\sqrt{\alpha'}g_s$, $\Theta=(H^{-1}-1)$, $\Gamma={1\over 5!\sqrt{-g}}\epsilon^{ijklm}\Gamma_{ijklm}\Gamma^{\underline \varphi}$, the moduli metric $g_{\dot m \dot n}=H^{1/2}\delta_{\dot m \dot n}$ and the worldvolume metric $g_{ij}$
\bea
g_{ij}=\left( \begin{array}{cc}
    -H^{-1/2} & 0 \\
    0 & H^{1/2}\delta_{IJ}
    \end{array}\right).
\eea

In order to display this theory as a standard field theory in a curved background, we now wish to fix $\kappa$-symmetry and identify the remaining supersymmetries. We choose the $\kappa$-symmetry gauge specified by the condition
\be
\label{kcond}
\bar y {1\over 2}(1-\Gamma)= \bar y
\ee 
That is, in terms of the
decomposition of the general 32 Majorana spinor $y$ into $y_+ + y_-$ where $\Gamma y_{\pm}=\pm y_{\pm}$, our gauge sets $y_-=0$. 

From now on we refer to the fermion $y$ satisfying (\ref{kcond}) as $\psi$.  In terms of $\psi$ the action simplifies slightly to
\begin{equation}
\label{kfS}
S^{(2)}_{D4}=\frac
{i}{g^2_{YM}}\int d^5\xi \   \bar \psi\left[\Gamma^i\partial_i-{1\over 8}\partial_I\ln H \Gamma^I(1+2\Gamma_{\underline0}\Gamma^{\underline \varphi})\right]
\psi  \ .
\end{equation}

The background killing spinor can also be decomposed into eigenspinors of $\Gamma$. It is then not difficult to see that only the supersymmetry transformation related to $\varepsilon_-$ are relevant in our gauge\footnote{The transformations generated by the other killing spinors $\varepsilon_+$, correspond only to the (position dependent) translations in field space  $\psi \rightarrow \psi + \varepsilon_+$.
The variation of $S_{D4}^{(2)}$ under any such translation vanishes due to the Killing equation satisfied by $\varepsilon_+$.  Since this 
transformation acts trivially on the bosonic fields, it has nothing to do with supersymmetry from the viewpoint of the $\kappa$-symmetry fixed worldvolume theory.}, and that the associated transformation rules are
\bea
\label{D4Susy}
&&\delta_{\varepsilon_-} \psi = \big(\hbox{${1\over4}$}F^{ij}\Gamma_{ij}\Gamma^{\underline \varphi}+\hbox{${1\over 2}$}\partial_i\Phi^{\dot m}\Gamma^i\Gamma_{\dot m}\big)\varepsilon_-\ , \cr
&&\delta_{\varepsilon_-} A_i = i\bar \varepsilon_-\Gamma_i\Gamma^{\underline \varphi}\psi \ ,\cr
&&\delta_{\varepsilon_-} \Phi^{\dot m} =  i\bar \varepsilon_-\Gamma^{\dot m}\psi \ ,
\eea
where $\varepsilon_-=H^{-1/8}\varepsilon^{(0)}_-$ and  $\varepsilon^{(0)}_-$ is a constant spinor that satisfies
\bea
&&(1+\Gamma_{\underline 0}\Gamma^{\underline \varphi})\varepsilon^{(0)}_-=0 \, \ \  \ {\rm and,}\cr
&&(1+\Gamma_{\underline 0 \underline 1 \underline 2 \underline 3 \underline 4}\Gamma^{\underline \varphi})\varepsilon^{(0)}_-=0
\eea
To obtain the linear transofmations (\ref{D4Susy}), we have combined the supersymmetry transformations (\ref{generalsusy}) with an appropriate 
$\kappa$-symmetry transformation so that the gauge condition (\ref{kcond}) is preserved.
Note that the two projectors commute, and therefore that 1/4 of the 32 supersymmetries survives.
Recall that in (\ref{D4Susy}) $i,j,\dot m$ represent spacetime indices either along ($ij$)
or transverse ($\dot m$) to the D4-brane.

The commutator of two supersymmetry transformations corresponding to $\varepsilon^{1}_-,\varepsilon^{2}_-$  acting on a bosonic field ($\Phi$ or $A$) is readily computed to be
\begin{equation}
[ \delta_{\varepsilon^{1}_-},
\delta_{\varepsilon^{2}_-} ] = \left( -i \bar \varepsilon^{2}_-  \Gamma^0 \varepsilon^{1}_- \right)  \partial_0 - Q\left[ i \bar \varepsilon^{2}_-  \Gamma^0 A_0 \varepsilon^{1}_- \right],
\end{equation}
where $Q$ is the generator of gauge transformations; i.e. $Q[\Lambda] \Phi = Q[\Lambda] \psi =0$, but $Q[\Lambda] A_i = \partial_i \Lambda$.  In reaching the above form we have used the fact that, since $\Gamma_{\underline 1234}
\varepsilon_- =\varepsilon_-$, one has $ -i \bar \varepsilon^{2}_-  \Gamma^I \varepsilon^{1}_-  =0$.
Note that the factors of $H$ in the first term cancel so that it represents a constant time translation, which is indeed a symmetry of the action (\ref{D4inD0}).  

\section{Summary}

In this article we have used normal coordinate expansions to
obtain the fully interacting actions to second order in
fermions for D$p$-branes in arbitrary bosonic type II supergravity
backgrounds. This completes the analysis begun in our previous
work \cite{mms1}, where certain interaction terms were suppressed.
The derivation of the D-brane actions was carried
out using two different (but not entirely independent) 
methods that produce two different-looking sets of
actions. However, after some algebraic manipulation these two sets proved
to be equivalent.  Thus we have provided a useful cross-check that further supports the validity of the expressions given in (\ref{Dpt}),(\ref{Born-Infeldaction}), and (\ref{sDp2}).

We have also discussed the supersymmetry and $\kappa$-symmetry properties of our actions, including that of the M2-brane. Note that before fixing $\kappa$-symmetry, we have as many supersymmetries as the background.  These act non-linearly in the form
displayed in section 4, but often some fraction can be realized linearly by acting simultaneously with an appropriate $\kappa$-symmetry
transformation.  This was seen explicitly in the D0/D4 example of section 5 in which the resulting theory was shown to be invariant under the action of 8 supercharges.

We expect the technology developed in this work to have a number
of applications.  The most important of these will likely involve
the study of D-branes in the supergravity backgrounds generated by
other D-branes.  For example, we saw in section 5 that this technology is ideally suited to
studying the D$p$/D(p+4) system.  This system is known to preserve
supersymmetry when the branes are separated so that one may consider a probe 
D$(p+4)$-brane located in a non-singular region of, e.g., the D$p$ supergravity solution (or vice-versa).
The case of the D4-brane in the background generated by a large number of D0-branes was analyzed in detail in
section 5, in particular, we found in the limit $\lambda \rightarrow 0$ a non-interacting curved spacetime field theory with linearly realized supersymmetry. In addition, the D-instanton background may be of particular interest in studying effects in
4-dimensional super Yang-Mills theory, though information about
the non-abelian case may need to await a generalization of our
results to multi-brane systems.

\vspace{1cm}
{\bf Acknowledgments}\\

We thank M. Grisaru, R. Myers and D. Zanon for useful discussions. L. Martucci and P. J. Silva were partially supported by INFN, MURST and by the European Commission RTN program HPRN-CT-2000-00131, in association with the University of Torino. D. Marolf and P. J. Silva were supported in part by NSF grant PHY00-98747 and by funds from Syracuse University.

\vspace{2cm}

\noindent{\LARGE\bf Appendices}

\appendix

\section{Spinor conventions and Gamma matrix algebra}\label{conventions}

This appendix is a list of spinor conventions.

In 11D superspace, we denote the general supercoordinates by $z^M$, where $M$ runs over the bosonic coordinates $x^m$, and fermionic coordinates $\theta^\mu$. Thus the curved index $M$ splits into $M=(m,\mu)$, where $m=0,1,...,10$, $\mu=1,2,...,32$. We use $A=(a,\alpha)$ for tangent space indices. We also underline explicit tangent space indices (e.g., {\ul 0}, {\ul1}, etc.), to differentiate them from explicit space-time indices.

We take the metric to have signature $(-,+,...,+)$ and use the Clifford algebra

\be
\{\Gamma^a,\Gamma^b\}=2\eta^{ab}\;,
\ee
where $\Gamma^a$ are real gamma matrices and $\eta^{ab}$ is the 11D Minkowski metric. We also set $\epsilon^{01\ldots}=1$ and use the notation
$\Gamma_{a_1...a_n}=\Gamma_{[a_1}...\Gamma_{a_n]}$
denoting antisymmetrization with weight one; e.g. $\Gamma_{01} = \hbox{${1\over2}$}(\Gamma_0 \Gamma_1- \Gamma_1 \Gamma_0) = \Gamma_0 \Gamma_1$.

We use real Majorana anticommuting spinors of 32 components, denoted $y^\alpha$ or $\theta^\mu$.
The conjugation operation is defined by,

\bea
\bar{y}=y^TC\;, \nn
\bar y_\beta = y^\alpha C_{\alpha\beta}\;,
\eea
where $T$ corresponds to transpose matrix multiplication; e.g. $y^\alpha C_{\alpha\beta}$ instead of $C_{\alpha\beta}y^\beta$, and $C=C_{\alpha\beta}$ is the antisymmetric charge conjugation matrix with inverse $C^{-1}=C^{\alpha\beta}$. The indices of a spinor and a bispinor $M^\alpha_{\;\;\beta}$ are lowered and raised via matrix multiplication by $C$ so that we have

\bea
C_{\alpha\beta}C^{\beta\gamma}=\delta^{\;\;\gamma}_{\alpha}\;, \nn
M_\alpha^{\;\;\beta}=C_{\alpha\gamma}M^\gamma_{\;\;\delta}C^{\delta\beta}\;, \nn
\bar{\theta}M\xi=\bar\theta_\alpha M^\alpha_{\;\;\beta} \xi^\beta =\theta^\alpha M_{\alpha\beta} \xi^\beta\;.
\eea
We take $C=\Gamma^{\ul{0}}$. It should also be noted that for Majorana spinors
like $y$, any expression $\bar{y}\Gamma_{a_1..a_n}y$ vanishes for $n=(1,2,5,6,9,10)$ but in general is non-vanishing for $n=(0,3,4,7,8)$. For Majorana-Weyl spinors, only the corresponding expressions for $n=(3,7)$ can be non-vanishing.

Once one of the directions, say $x^{10}$, has been compactified and the corresponding
$\Gamma^{\ul{10}}$ is identified with the chiral gamma matrix $\Gamma^{\ul{\varphi}}$,
the fermionic coordinates appearing in 11D supergravity can be decomposed into two Majorana
Weyl spinors (each of which we write in 32-component form). Thus in
type IIA we may write

\be
y=y_++y_-\;\;\hbox{where}\;\;\Gamma^{\ul{\varphi}} y_\pm=\pm y_\pm\;.
\ee
In type IIB, we choose the two 32 real component chiral spinors $y_1,y_2$ to have positive
chirality, and we write them together as a 64-component spinor of the form

\bea
y=\left(
\begin{array}{cc}
y_1\\
y_2\\
\end{array}\right).
\eea
Taking the tensor products of the 32 $\times$ 32 component $\Gamma^a$ and $\Gamma^{\ul{\varphi}} $matrices
with respectively the 2 $\times$ 2 identity operator and $\sigma_3$ yields the 64 $\times$ 64 matrices

\bea
&&\Gamma^a = \left( \begin{array}{cc}
\Gamma^a & 0 \\
0  & \Gamma^a \\
\end{array}\right)\ , \  \hat \Gamma^{\ul{\varphi}} = \left( \begin{array}{cc}
\Gamma^{\ul{\varphi}} & 0 \\
0  & -\Gamma^{\ul{\varphi}} \\
\end{array}\right)\ . \nonumber
\eea
Finally, we use the usual Pauli matrices,
\[
\sigma_1 = \left( \ba{cc} 0 & 1 \\ 1 & 0 \ea \right)
\qquad \qquad \sigma_2 = \left( \ba{cc} 0 & -i \\ i & 0 \ea \right)
\qquad \qquad \sigma_3 = \left( \ba{cc} 1 & 0 \\ 0 & -1 \ea \right) \]


\section{Supergravity} \label{sugra}

This appendix is a list of supergravity conventions. We highlight that in this paper we always use the superspace convention for differential form; i.e.,
\bea\label{forms}
w^{(p)}=\hbox{${1\over p!}$}dx^{m_1}\wedge\cdots \wedge dx^{m_p} w_{m_p\cdots m_1}\ .
\eea

\subsection{11D supergravity}

Here we borrow some conventions and definitions directely from Grisaru and Knutt \cite{gk1}. We also use, in the main body of the paper, bold letters for the pull-backs to the brane of bulk superfields.

The theory is described in terms of the vielbein ${E}^A (x, \theta) = dZ^M {E_M}^A$ and three-form $A =(1/3!)E^CE^BE^A A_{ABC}$  satisfying respectively the torsion and field-strength constraints\cite{Howe, Cremmer}:

\bea
&&{T_{\a \b}}^c = -i (\Gamma^c)_{\a \b} \nonumber\\
&&{T_{\a \b}}^{\gamma} = {T_{\a b}}^c ={T_{a b}}^c=0\nonumber\\
&&H_{\a \b \gamma \d} = H_{\a \b \gamma d}=H_{\a bcd}=0 \nonumber\\
&&H_{\a \b cd}= i (\Gamma_{cd})_{\a \b}
\eea
with $H = dA =(1/4!)E^DE^CE^BE^AH_{ABCD}$ and

\be
H_{ABCD}= \sum_{(ABCD)}\nabla_A A_{BCD} +{T_{AB}}^E A_{ECD}.
\ee
These constraints put the theory on shell. From the Bianchi identities $DT^A = E^B{R_B}^A$, $D{R_A}^B=0$ and
$dH=0$, or one derives \cite{Howe} expressions for the remaining components of the torsion:
\bea
\label{torsolutions}
&&{T_{a \b}} ^\gamma= \frac{1}{36}{(\d_a^b\Gamma^{cde}+\frac{1}{8}{\Gamma_a}^{bcde})_{\b}}^\gamma H_{bcde}
\nonumber\\
&&{T_{ab}}^\a=\frac{i}{42}{(\Gamma^{cd})}^{\a\b} \nabla_\b H_{abcd}\\
&&(\Gamma^{abc})_{\a\b}{T_{bc}}^ \b=0\nonumber
\eea
From the constraints and the Bianchi identities, one obtains the usual 11d supergravity, whose bosonic 
part of the action is
\bea
\label{11sugra}
 S_{11d} &=&
\frac{1}{2\kappa_{11}^2}\int d^{11} x \sqrt{-g}
    \Big(R  -\frac{1}{2 \cdot 4!} H^2\Big) -  \frac{1}{12\kappa_{11}^2}\int A\wedge H\wedge H\ ,
\eea
where $2\kappa_{11}^2=(2\pi)^8 l_p^9$, with $l_p$ the 11d Plank length.
The supersymmetry transformation low for the 11d gravitino is $\delta_\varepsilon \psi_m=\tilde D_m \varepsilon$, where
\bea\label{gravitino11d}
\tilde D_m=\nabla_m -\frac{1}{288}(\Gamma_m{}^{pqrs}-8\delta_m^p\Gamma^{qrs})H_{pqrs}\ .
\eea
.

\subsection{10D supergravity}

Here we use essentially the same conventions as in \cite{has,antoine}\footnote{The results of \cite{has}
are well suited for our convention (\ref{forms}) for differential forms. This translates in a particular 
choice of the overall sign of the RR fields. For example, the link with the fields  of \cite{myers} 
can be obtained by the substitution $C^{(n)}_{m_1\ldots m_n}\rightarrow (-)^{\frac{n(n-1)}{2}}
C^{(n)}_{m_1\ldots m_n}$.}.  

First, we note that two types of RR field strength appear in the literature of type II supergravity. In addition to $dC^{(n)}$, it is also useful to introduce:

\begin{eqnarray}
{\bf F}^{(1)} &=& dC^{(0)}\cr
{\bf F}^{(2)} &=& dC^{(1)} \cr
{\bf F}^{(3)} &=& dC^{(2)} -C^{(0)}\,H\cr
{\bf F}^{(4)} &=& dC^{(3)} -C^{(1)}\wedge H\cr
{\bf F}^{(5)} &=& dC^{(4)} -C^{(2)}\wedge H\ .
\end{eqnarray}

The type IIA bosonic part of the action is given by

\bea
 S_{IIA} &=&
\frac{1}{2\kappa_{10}^2}\int d^{10} x \sqrt{-g}
    \Big\{
    e^{-2\phi} \big[
    R +4\big( \partial{\phi} \big)^{2}
    -\frac{1}{2 \cdot3!} H^2\big] + \nn
   && - \frac{1}{2\cdot 2!} ({\bf F}^{(2)})^2 - \frac{1}{2\cdot 4!} ({\bf F}^{(4)})^2 \Big\}
+ \frac{1}{4\kappa_{10}^2}\int b\wedge dC^{(3)}\wedge dC^{(3)}\ , \cr &&
\eea
and the supersymmetry transformations for the gravitino $\psi_m$ and dilatino $\lambda$ are,
\bea
\delta\psi_m &=& \left[\partial_m +\frac{1}{4} \omega_{mab}\Gamma^{ab}+\frac{1}{4\cdot 2!}H_{mab}\Gamma^{ab}\Gamma^{\ul{\varphi}} \; + \right. \nn
&&\left. + \frac18 e^\phi \big( \frac{1}{2!} {\bf F}^{(2)}_{ab}\Gamma^{ab}\Gamma_m\Gamma^{\ul{\varphi}}+ \frac{1}{4!}{\bf F}^{(4)}_{abcd}\Gamma^{abcd}\Gamma_m\big)\right]\epsilon \ , \nn
\delta\lambda &=& \left[ \frac12 \left( \Gamma^m \partial_m\phi + \frac{1}{2\cdot 3!}H_{abc}\Gamma^{abc}\Gamma^{\ul{\varphi}}\right) \; +\right.\nn
&&+ \left. \frac{1}{8} e^\phi \left( \frac{3}{2!} {\bf F}^{(2)}_{ab}\Gamma^{ab}\Gamma^{\ul{\varphi}}+ \frac{1}{4!} {\bf F}^{(4)}_{abcd}\Gamma^{abcd}\right)\right] \epsilon \ .
\end{eqnarray}

The type IIB bosonic part of the action is given by\footnote{In \cite{mms1,ms1} a term was forgotten in the Chern-Simons part of the type IIB supergravity lagrangian.}
\bea
S_{IIB}&& =
\frac{1}{2\kappa_{10}^2}\int d^{10} x \sqrt{-g}
    \Big\{
    e^{-2\phi} \big[
    R +4\big( \partial{\phi} \big)^{2}
    -\frac{1}{2 \cdot3!} H^2\big] + \nn
   && - \frac{1}{2} ({\bf F}^{(1)})^2 - \frac{1}{2\cdot 3!} ({\bf F}^{(3)})^2
- \frac{1}{4\cdot 5!} ({\bf F}^{(5)})^2  \Big\}
+\cr
&&+\frac{1}{4\kappa_{10}^2}\int db\wedge dC^{(2)}\wedge\Big(C^{(4)}-\frac12 b\wedge C^{(2)}\Big)\ ,
\eea
and the supersymmetry transformations for the gravitino $\psi_m$ and dilatino $\lambda$ are,

\bea
\delta\psi_{(1,2)m} &=& \left(\partial_m +\frac{1}{4} \omega_{mab}\Gamma^{ab}\pm\frac{1}{4\cdot 2!}H_{mab}\Gamma^{ab}\right)\epsilon_{(1,2)} +  \cr
&&+ \frac18 e^\phi \left(\pm {\bf F}^{(1)}_a\Gamma^a - \frac{1}{3!} {\bf F}^{(3)}_{abc}\Gamma^{abc}\pm
\frac{1}{2\cdot 5!}{\bf F}^{(5)}_{abcde}\Gamma^{abcde}\right)\Gamma_m \epsilon_{(2,1)}\;, \nn
\delta\lambda_{(1,2)} &=& \frac12 \left( \Gamma^m \partial_m\phi \pm\frac{1}{2\cdot 3!}H_{abc}\Gamma^{abc}\right)\epsilon_{(1,2)}\;+ \nn 
&&\hspace{2,6cm}+ \frac{1}{2} e^\phi \left( \mp  {\bf F}^{(1)}_{a}\Gamma^{a}+
\frac{1}{2\cdot 3!} {\bf F}^{(3)}_{abc}\Gamma^{abc}\right) \epsilon_{(2,1)}\; .
\eea

In the above expressions $2\kappa_{10}^2=(2\pi)^7 l_s^8 g_s^2$ and for the type IIB case we use the convention that the self duality constraint on ${\bf F}^{(5)}$ is imposed by hand at the level of the equations of motion.


\section{T-duality rules}\label{T-duality}
We perform T-dualities using the Hassan formalism\footnote{In \cite{ms1} this formalism was shown to be consistent with the T-duality relation between type IIa and type IIb superstrings.} \cite{has}.
In this approach, the T-duality transformations are closely related to the supersymmetry transformations of the gra\-vitino ($\delta \psi_m\sim D_m \epsilon$ ) and the dilatino ($\delta\lambda\sim \Delta \epsilon$). These supersymmetry transformations involve the operators acting on 10D Majorana spinors in type IIA  supergravity (they have been just introduced in section (\ref{D2section})):
\begin{eqnarray}\label{operatorsIIA}
D_m &=& D^{(0)}_m+W_m \cr
\Delta &=& \Delta^{(1)}+\Delta^{(2)}\ ,
\end{eqnarray}
where
\begin{eqnarray}
D^{(0)}_m &=& \partial_m +\frac{1}{4} \omega_{mab}\Gamma^{ab}+\frac{1}{4\cdot 2!}H_{mab}\Gamma^{ab}\Gamma^{\ul{\varphi}} \cr
W_m &=& \frac18 e^\phi \left( \frac{1}{2!} {\bf F}^{(2)}_{ab}\Gamma^{ab}\Gamma_m\Gamma^{\ul{\varphi}}+
\frac{1}{4!}{\bf F}^{(4)}_{abcd}\Gamma^{abcd}\Gamma_m\right)\cr
\Delta^{(1)} &=& \frac12 \left( \Gamma^m \partial_m\phi +\frac{1}{2\cdot 3!}H_{abc}\Gamma^{abc}\Gamma^{\ul{\varphi}}\right)\cr
\Delta^{(2)}&=& \frac{1}{8} e^\phi \left( \frac{3}{2!} {\bf F}^{(2)}_{ab}\Gamma^{ab}\Gamma^{\ul{\varphi}}+
\frac{1}{4!} {\bf F}^{(4)}_{abcd}\Gamma^{abcd}\right)\ .
\end{eqnarray}
It is also convenient to decompose the Majorana spinors in terms of
the Weyl spinors of type IIA and IIB, hence we split our Majorana spinor
$y$ into two Majorana-Weyl (MW) spinors of opposite chirality:
\begin{eqnarray}
y=y_+ + y_-\ ,\ \Gamma^{\ul{\varphi}}\;y_\pm=\pm \; y_\pm\ .
\end{eqnarray}
Therefore, when acting on MW spinors of
chirality $\pm$, the operators above take the form
\begin{eqnarray}
D_{(\pm)m} &=& D^{(0)}_{(\pm)m}+W_{(\pm)m},\cr
\Delta_{(\pm)} &=& \Delta^{(1)}_{(\pm)}+\Delta^{(2)}_{(\pm)}\ ,
\end{eqnarray}
with
\begin{eqnarray}
D^{(0)}_{(\pm)m} &=& \partial_m +\frac{1}{4} \omega_{mab}\Gamma^{ab}\pm \frac{1}{4\cdot 2!}H_{mab}\Gamma^{ab}\cr
W_{(\pm) m} &=& \frac18 e^\phi \left(\pm \frac{1}{2!} {\bf F}^{(2)}_{ab}\Gamma^{ab}+
\frac{1}{4!}{\bf F}^{(4)}_{abcd}\Gamma^{abcd}\right)\Gamma_m\cr
\Delta^{(1)}_{(\pm)} &=& \frac12 \left( \Gamma^m \partial_m\phi \pm\frac{1}{2\cdot 3!}H_{abc}\Gamma^{abc}\right)\cr
\Delta^{(2)}_{(\pm)}&=& \frac{1}{8} e^\phi \left( \pm\frac{3}{2!} {\bf F}^{(2)}_{ab}\Gamma^{ab}+
\frac{1}{4!} {\bf F}^{(4)}_{abcd}\Gamma^{abcd}\right)\ .
\end{eqnarray}
In the rest of this work, we will not write the subscript $(\pm)$
explicitly, as it will be determined by the chirality of
the spinor on which the operators act.

For type IIB supergravity theory, we have two MW spinors $y_{(1,2)}$
of positive chirality and the following operators acting on them
(the upper sign refers to $y_1$ while the lower one to $y_2$):
\begin{eqnarray}
\hat D^{(0)}_{(1,2)m} &=& \partial_m +\frac{1}{4} \omega_{mab}\Gamma^{ab}\pm \frac{1}{4\cdot 2!}H_{mab}\Gamma^{ab}\cr
\hat W_{(1,2) m} &=& \frac18 e^\phi \left(\mp {\bf F}^{(1)}_a\Gamma^a - \frac{1}{3!} {\bf F}^{(3)}_{abc}\Gamma^{abc}\mp
\frac{1}{2\cdot 5!}{\bf F}^{(5)}_{abcde}\Gamma^{abcde}\right)\Gamma_m\cr
\hat\Delta^{(1)}_{(1,2)} &=& \frac12 \left( \Gamma^m \partial_m\phi \pm\frac{1}{2\cdot 3!}H_{abc}\Gamma^{abc}\right)\cr
\hat\Delta^{(2)}_{(1,2)}&=& \frac{1}{2} e^\phi \left( \pm  {\bf F}^{(1)}_{a}\Gamma^{a}+
\frac{1}{2\cdot 3!} {\bf F}^{(3)}_{abc}\Gamma^{abc}\right)\ .
\end{eqnarray}
As for the type IIA operators, we will suppress the subscript
$(1,2)$.  This subscript is determined by the spinor on which the operators act.
In the paper we usually use the double spinor convention introduced in appendix \ref{conventions} for type IIB backgrounds. With this notation,it is useful to introduce the analog of operators (\ref{operatorsIIA}) also for the type IIB case (we use the same symbol for both the case):
\bea
\label{operatorsIIB}
\hat D_m &=& \hat D^{(0)}_m+\sigma_1 \otimes \hat W_m \cr
\hat \Delta &=& \hat\Delta^{(1)}+\sigma_1\otimes\hat\Delta^{(2)}\ . 
\eea
We wish to apply T-duality along the 9th direction. Let us introduce
the following useful objects ($\hat m,\hat n=0,\ldots,8$)
\begin{eqnarray}
&& \Omega=\frac{1}{\sqrt{g_{99}}}\Gamma^{\ul{\varphi}}\Gamma_9 \Rightarrow \Omega^2=-1\cr
&& E_{mn}=g_{mn}+b_{mn}\cr
&& (Q_{\pm})^m{}_n=\left(
\begin{array}{cc}
\mp g_{99} & \mp (g\mp b)_{9\hat n}\\
 0         &   {\bf 1}_9 \\
\end{array} \right)\cr
&& (Q^{-1}_{\pm})^m{}_n=\left(
\begin{array}{cc}
\mp g_{99}^{-1} & -g_{99}^{-1}(g\mp b)_{9\hat n}\\
 0         &   {\bf 1}_9 \\
\end{array} \right)\ .
\end{eqnarray}
The T-duality rules for $E_{mn}$ are\footnote{Here and in the rest of this work,
we place a tilde over the transformed fields to remove ambiguity when needed.},
\begin{eqnarray}
\tilde \phi &=& \phi -\frac 12 ln\; g_{99}\cr
\tilde E _{\hat m\hat n}&=& E _{\hat m\hat n}-E _{\hat m 9}g_{99}^{-1}E _{9\hat n}\cr
\tilde E _{\hat m 9}&=& E _{\hat m 9}g_{99}^{-1}\cr
\tilde E _{9 \hat m }&=& -E _{9\hat m }g_{99}^{-1}\cr
\tilde E _{9 9 } &=& g_{99}^{-1}\ .
\end{eqnarray}
For the transformation of the vielbein and the spinors, we will use the Hassan
conventions to avoid ambiguities\footnote{Recall that there are two possible choices $e^m_{(\pm)a}$ for the transformed vielbein, related by a Lorentz transformation $\Lambda^b{}_{a}$.}.
The transformation rules for the vielbein are
\begin{eqnarray}
\tilde e^m_a\equiv e^m_{(-)a}= (Q_{-})^m{}_n e ^n_a \Rightarrow \tilde e ^a_m\equiv e^a_{(-)m}= (Q_{-}^{-1})^n{}_m e ^a_n\ .
\end{eqnarray}
We will also need the alternative transformed vielbein
\begin{eqnarray}
 e^m_{(+)a}= (Q_{+})^m{}_n e ^n_a=\Lambda^b{}_{a} e^m_{(-)b} \Rightarrow e^a_{(+)m}=
(Q_{+}^{-1})^n{}_m e ^a_n=e^b_{(-)m}\Lambda_b{}^a\ .
\end{eqnarray}
At last we present the transformation rules for the RR potentials $C^{(n)}$
\bea
\tilde C^{(n)}_{9\hat m_2\cdots \hat m_n}&=&C^{(n-1)}_{\hat m_2\cdots \hat m_n}-(n-1)g^{-1}_{99}g_{9[\hat m_2}C^{(n-1)}_{|9|\hat m_3\cdots \hat m_n]}\;, \nn
\tilde C^{(n)}_{\hat m_1\cdots \hat m_n}&=&C^{(n+1)}_{9\hat m_1\cdots \hat m_n}-nb_{9[\hat m_1}\tilde C^{(n)}_{|9|\hat m_2\cdots \hat m_n]}\;.
\label{trr2}
\eea
Therefore going from IIA to IIB, we have:
\begin{eqnarray}
y_+ &=& y_1 \Rightarrow  \bar y_+ = \bar y_1\cr
y_- &=& -\Omega y_2 \Rightarrow  \bar y_- = \bar y_2 \Omega\cr
&&\cr
D^{(0)}_m y_+ &=& (Q_{+}^{-1})^n{}_m(\hat D^{(0)}_n y_1)\cr
D^{(0)}_m y_- &=& -\Omega(Q_{-}^{-1})^n{}_m(\hat D^{(0)}_n y_2)\cr
W_m y_+ &=& -\Omega(Q_{-}^{-1})^n{}_m(\hat W_n y_1)\cr
W_m y_- &=& (Q_{+}^{-1})^n{}_m(\hat W_n y_2)\cr
&&\cr
\Delta^{(1)}y_+ &=& \hat \Delta^{(1)}y_1 -g_{99}^{-1}\Gamma_9 \hat D ^{(0)}_9 y_1\cr
\Delta^{(1)}y_- &=& -\Omega(\hat \Delta^{(1)}y_2 -g_{99}^{-1}\Gamma_9 \hat D ^{(0)}_9 y_2)\cr
\Delta^{(2)}y_+ &=& -\Omega(\hat \Delta^{(2)}y_1 -g_{99}^{-1}\Gamma_9 \hat W_9 y_1)\cr
\Delta^{(2)}y_- &=& \hat \Delta^{(2)}y_2 -g_{99}^{-1}\Gamma_9 \hat W_9 y_2\ .
\end{eqnarray}
Conversely, going from IIB to IIA we have
\begin{eqnarray}
y_1 &=& y_+ \Rightarrow  \bar y_1 = \bar y_+\cr
y_2 &=& \Omega y_- \Rightarrow  \bar y_2 = -\bar y_- \Omega\cr
&&\cr
\hat D^{(0)}_m y_1 &=& (Q_{+}^{-1})^n{}_m (D^{(0)}_n y_+)\cr
\hat D^{(0)}_m y_2 &=& \Omega(Q_{-}^{-1})^n{}_m(D^{(0)}_n y_-)\cr
\hat W_m y_1 &=& \Omega(Q_{-}^{-1})^n{}_m(W_n y_+)\cr
\hat W_m y_- &=& (Q_{+}^{-1})^n{}_m(W_n y_-)\cr
&&\cr
\hat\Delta^{(1)}y_1 &=& \Delta^{(1)}y_+ -g_{99}^{-1}\Gamma_9 D ^{(0)}_9 y_+\cr
\hat\Delta^{(1)}y_2 &=& \Omega(\Delta^{(1)}y_- -g_{99}^{-1}\Gamma_9 D ^{(0)}_9 y_-)\cr
\hat\Delta^{(2)}y_1 &=& \Omega(\hat \Delta^{(2)}y_+ -g_{99}^{-1}\Gamma_9  W_9 y_+)\cr
\hat \Delta^{(2)}y_2 &=& \Delta^{(2)}y_- -g_{99}^{-1}\Gamma_9 W_9 y_-\ .
\end{eqnarray}

\subsection{T-duality to the D2-brane action}

Here we obtain the D$p$-brane actions for all $p$ by T-duality from the D2-brane. Instead of a completely brute force
calculation, we use a short-cut that relies on the general structure of the D$p$-brane actions. Note that the analysis of \cite{mms1} shows that T-duality on (\ref{part1}) yields a result of the form
\bea
\label{part2}
\frac {iT_{Dp}}{2}\int d^{p+1}\xi
e^{-\phi}\sqrt{-(g+{\cal F})}\bar y (1-\tilde\Gamma_{Dp})\big[(\Gamma^iD_i-\Delta)y + O({\cal F})\big]\ .
\eea
We will obtain the $O(\cal F)$ terms by means of a trick, rather than by brute force. 

First recall from section \ref{superactions} that the D$p$-brane actions have a Born-Infield part and a Chern-Simons part. 
These are characterized as follows:  the Born-Infeld part is invariant under reversing the orientation of the brane while the Chern-Simons part changes sign.  As a result, any term with a definite transformation under orientation reversal can be
identified as a Born-Infeld or Chern-Simons term\footnote{In practice, this amounts to counting the number of Levi-Civita
tensors in the term.}.  Since reversing orientation corresponds to changing the sign of the D-brane charge, it is clear that this characterization is preserved under T-duality.

Let us now return to the D2-brane action (\ref{d2action2}) of section 3.
Using (\ref{sfields}), it is not dificult to isolate the corresponding Chern-Simons terms quadratic in fermions from (\ref{d2action2}).  They are \bea 
\label{superWZ}
S^{Chern-Simons(2)}_{D2}&=&-\frac{iT_{D2}}{2\cdot 3!}\int d^3\xi
e^{-\phi}\epsilon^{ijk}\bar y\big[ 3\Gamma_{ij}D_k -
\Gamma_{ijk}\Delta\big]y+\cr
&&- \frac {iT_{D2}}{4} \int d^3\xi
e^{-\phi}\epsilon^{ijk}{\cal F}_{kj}\bar y
\Gamma^{\ul{\varphi}}(D_i-\Gamma_i\Delta) y \ .
\eea 
Using the results listed earlier in this appendix, it is then straightforward to T-dualize each term to the D3-brane, obtaining 
\bea
S^{Chern-Simons(2)}_{D3}&=&T_{D3}\int \left( {\bf C}^{(4)}_{(2)}- {\bf C}^{(2)}_{(2)}\wedge{\cal F}+{1\over2}{\bf C}^{(0)}_{(2)}{\cal F}\wedge{\cal F}\right) \ ,
\label{cs3}
\eea  
where ${\bf C}^{(2n)}_{(2)}$ stands for the part of the potentials defined in (\ref{srrp}) which are quadratic in fermions. Note that we have obtained all of the terms of the Chern-Simons part of the D3-brane\footnote{We can therefore perform further T-dualities to obtain the Chern-Simons parts of the actions of any D$p$-brane.} action (\ref{sDp2} ).  This is as expected, since we have already observed that Chern-Simons terms
T-dualize to Chern-Simons terms.

Let us now compare (\ref{cs3}) with the action (\ref{part2}) for $p=3$.  Using the explicit form of $\tilde\Gamma_{D3}$, we see that expanding the term
\bea
\label{96}
-\frac {iT_{D3}}{2}\int d^{4}\xi
e^{-\phi}\sqrt{-(g+{\cal F})}\bar y \tilde\Gamma_{D3}(\Gamma^iD_i y-\Delta y)
\eea
yields the Chern-Simons terms of the D3-brane (\ref{cs3}) as well as the following extra terms 
\bea 
\label{extraC2}
&&-\frac {iT_{D3}}{2}\int d^4\xi
e^{-\phi}\sqrt{-(g+{\cal F})}\times \cr
&&\bar y \sum_{q\geq 1,q+r=2}\hbox{$\frac{\epsilon^{i_1\ldots i_{2q}j_1\ldots
j_{2r}}(-i\sigma_2)(\hGphi)^{r}}{q!2^q(2r)!\sqrt{-(g+{\cal
F})}}{\cal F}_{i_1i_2}\cdots{\cal
F}_{i_{2q-1}i_{2q}}\Gamma_{j_1\ldots j_{2r}}{}^k D_k y$}\ ;
\eea
that is (\ref{96}) is the sum of (\ref{cs3}) and (\ref{extraC2}).
Since no further Chern-Simons terms are explicitly displayed in (\ref{part2}), for $p=3$
the remaning terms of order $O({\cal F})$ in (\ref{part2})
must be such that when multiplied by $(1-\tilde\Gamma_{D3})$the results 
cancel (\ref{extraC2}) and yield no other terms of Chern-Simons form.

One immediately sees that this condition is satisfied if we take $O({\cal F})$ to be given by (\ref{dpactions2}) with
\bea
\label{counter3}
L_{3}=\sum_{q\geq 1,q+r=2}\hbox{$\frac{\epsilon^{i_1\ldots i_{2q}j_1\ldots j_{2r}}(-i\sigma_2)(\hGphi)^{r}}{q!2^q(2r)!\sqrt{-(g+{\cal F})}}{\cal F}_{i_1i_2}\cdots{\cal F}_{i_{2q-1}i_{2q}}\Gamma_{j_1\ldots j_{2r}}{}^k D_k$} .
\eea
Addition of further non-trivial terms is impossible as, when multiplied by $(1-\tilde\Gamma_{D_3})$, they would 
necessarily give rise to {\it both} Born-Infeld and Chern-Simons terms, while further Chern-Simons terms have already been excluded. Thus we have obtained the full set of $O({\cal F})$ terms and the full D3-brane action to second order in fermions.

Carrying out the above reasoning for the other branes one finds in general that the $O({\cal F})$ terms are given by (\ref{dpactions2}), with    
\bea
\label{counter}
L_{2n+1}&&=\sum_{q\geq 1,q+r=n+1}\hbox{$\frac{\epsilon^{i_1\ldots i_{2q}j_1\ldots j_{2r}}(-i\sigma_2)(\hGphi)^{r}}{q!2^q(2r)!\sqrt{-(g+{\cal F})}}{\cal F}_{i_1i_2}\cdots{\cal F}_{i_{2q-1}i_{2q}}\Gamma_{j_1\ldots j_{2r}}{}^k D_k$} \ ,\cr
L_{2n}=&&\sum_{q\geq 1,q+r=n}\hbox{$\frac{\epsilon^{i_1\ldots i_{2q}j_1\ldots
    j_{2r+1}}(-\Gphi)^{r+1}}{q!2^q(2r+1)!\sqrt{-(g+{\cal F})}}$}\hbox{${\cal
  F}_{i_1i_2}\cdots{\cal F}_{i_{2q-1}i_{2q}}\Gamma_{j_1\ldots j_{2r+1}}{}^k D_k$}\ , \cr &&
\eea
for D($2n$)- and D($2n+1$)-branes respectively. 

In summary, by combining the calculations of \cite{mms1} at zero order in ${\cal F}$, the general structure noted in section 2, and calculations to all orders in ${\cal F}$  for $\tilde\Gamma_{Dp}$ and the Chern-Simons term, we have obtained the unique completion of the fermionic part of the action (\ref{part2}).



\end{document}